\documentclass[prl,aps,twocolumn,showpacs,superscriptaddress]{revtex4}
\usepackage{bm,float,amssymb}
\usepackage[]{graphicx}
\usepackage{amsmath}

\begin{document}

\newcommand{\latin}[1]{\emph{#1}}
\newcommand{\etal}{\latin{et al.}}
\newcommand{\eg}{\latin{e.\,g.}}

\newcommand{\myav}[1]{\langle #1\rangle}
\newcommand{\uav}{\myav{u}}
\newcommand{\myvec}[1]{{\mathbf #1}}
\newcommand{\rvec}{\myvec{r}}
\newcommand{\tvec}{\myvec{t}}

\newcommand{\rhobar}{{\overline\rho}}

\title{The Shape of a Ponytail and the Statistical Physics of Hair
  Fiber Bundles}

\author{Raymond E. Goldstein}
\affiliation{Department of Applied Mathematics and Theoretical
Physics, University of Cambridge, Wilberforce Road, Cambridge CB3 0WA, UK.}

\author{Patrick B. Warren} \affiliation{Unilever R\&D Port Sunlight,
  Quarry Road East, Bebington, Wirral, CH63 3JW, UK.}

\author{Robin C. Ball} \affiliation{Department of Physics, University
  of Warwick, Coventry, CV4 7AL, UK.}

\date{\today}

\begin{abstract}
 
A general continuum theory for the distribution of hairs in a bundle
is developed, treating individual fibers as elastic filaments with
random intrinsic curvatures.  Applying this formalism to the iconic
problem of the ponytail, the combined effects of bending elasticity,
gravity, and orientational disorder are recast as a differential
equation for the envelope of the bundle, in which the compressibility
enters through an `equation of state'.  From this, we identify the
balance of forces in various regions of the ponytail, extract a
remarkably simple equation of state from laboratory measurements of
human ponytails, and relate the pressure to the measured random
curvatures of individual hairs.

\end{abstract}

\pacs{87.19.R-, 05.45.--a, 46.65.+g}

% 87.19.R- = Mechanical and electrical properties of tissues and organs 
% 05.45.-a = Fluctuation phenomena, random processes, noise, and Brownian motion
% 46.65.+g = Random phenomena and media (in continuum mechanics of solids)

\maketitle

\begin{figure}[b]
\begin{center}
\includegraphics[clip=true,width=0.90\columnwidth]{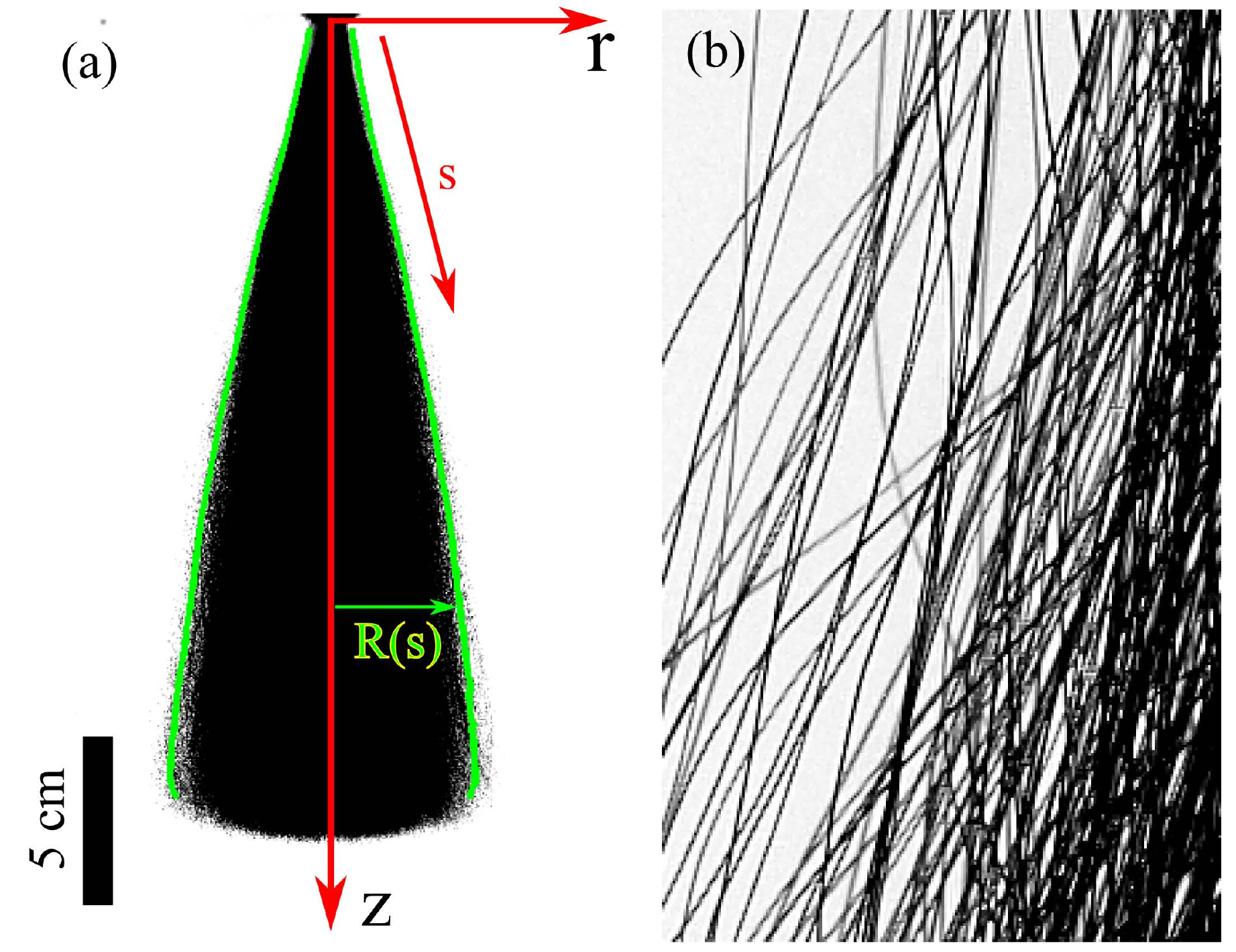}
\end{center}
\vskip -0.5cm
\caption{(color online) A ponytail. (a) Rotationally-averaged image of
  a switch of $N\approx 9500$ fibers, approximately $25$\,cm long.
  Coordinate system for envelope shape $R(s)$ in terms of arc length
  $s(z)$.  (b) Meanderings of hairs near ponytail edge.}
\label{fig1}
\end{figure}

One of the most familiar features of a bundle of hair such as a
ponytail is its `body' or `volume.'  Close examination reveals that
this property arises in a subtle way from the stiffness and shapes of
the individual fibers, whose meandering paths through the bundle
produce many collisions with other hairs (Fig.~\ref{fig1}).  These
meanderings are in part a consequence of the contacts themselves, but
hairs also have an \emph{intrinsic} waviness or curliness
\cite{Robbins,AudolyPomeau}.  Such curvatures may be generated during
growth, and vary with ethnicity.  They are clearly also modified by
chemical, thermal, and mechanical forces, as in the `water wave'
treatment, or a `perm' \cite{Robbins}.

From Leonardo to the Brothers Grimm our fascination with hair has
endured in art and science \cite{Leonardo,Hadap}. Yet, we still do not
have an answer to perhaps the simplest question that captures the
competing effects of filament elasticity, gravity, and mutual
interactions: {\it What is the shape of a ponytail?}  Note that the
average human has $\sim 10^5$ head hairs, so if even a modest fraction
is gathered into a ponytail, the number involved is enormous: this is
a problem in statistical physics.

\begin{figure*}
\begin{center}
\includegraphics*[clip=true,width=1.70\columnwidth]{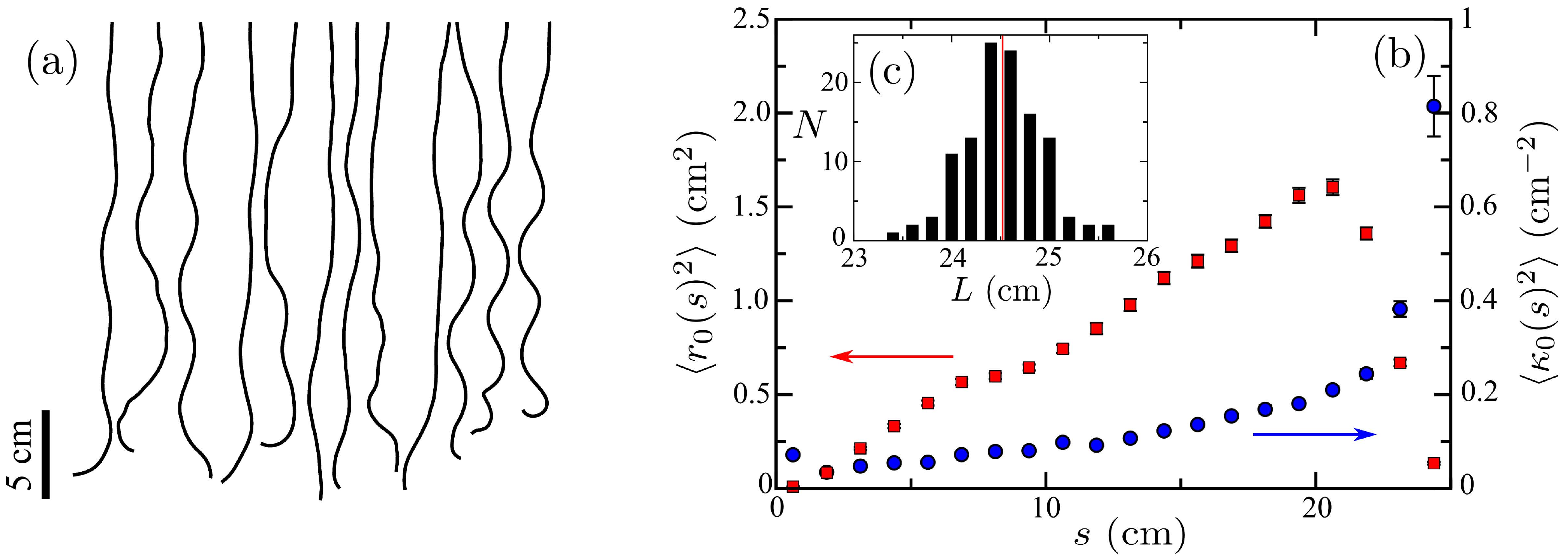}
\end{center}
\vskip -0.5cm
\caption{(color online) Geometry of hairs. (a) Representative
  projections of hair contours, thickened for clarity. (b) Mean
  squared radial excursion and curvature as functions of arc length,
  from processing stereoscopic image pairs \cite{details}.  The
  reconstructed arc lengths cluster tightly around $24.50\pm0.05$\,cm
  (inset histogram), demonstrating the accuracy of the image
  processing and analysis methodology.  Error bars are standard errors
  from ensemble averaging ($N=115$ fibers in total).}
\label{fig2}
\end{figure*}

Here we propose a theory for the ponytail shape on the basis of a
continuum theory for the spatial distribution of hairs in a bundle.
Their random curvatures give rise to a swelling pressure characterized
by an `equation of state' (EOS) of hair, a concept first introduced
semi-empirically by van Wyk in 1946 in relation to the compressibility
of wool \cite{vanWyk,Carnaby}, and explored recently for
two-dimensional randomly-curved fibers by Beckrich
\etal\ \cite{Beckrich}.  We exploit the predominantly vertical
alignment of hairs and axisymmetry to justify a number of
approximations that render the problem analytically tractable, and
thereby reduce the many-body problem to a one-body problem for the
ponytail {\it envelope}.  This shows how the EOS modifies the envelope
shape from that of a single hair bent by gravity, a classic problem in
elasticity \cite{AudolyPomeau}.

In parallel with the theoretical development, we measured the shapes
of ponytails made from commercially available hair `switches'
\cite{IHIP}, and of their component fibers.  Typical human hair has an
elliptical cross section and a distribution of major axis diameters
$40\lesssim d \lesssim 140$ $\mu$m.  We found $d = 79 \pm 16$\,$\mu$m
for a random sample from the switches.  Hair has an average density of
$\simeq 1.3$\,g/cm$^3$ \cite{Robbins}, and a linear mass density
$\lambda\simeq 65$\,$\mu$g/cm (in more amusing units, $6.5$ g/km).
Though its internal microstructure is complex, the bend and twist
moduli of hairs \cite{Robbins} are consistent with those of a
homogeneous incompressible material with a nylon-like modulus $E
\approx 4$\,GPa.  On the centimeter scale classical filament
elasticity holds, with a bending modulus $A = E \pi d^4 \!/ 64 \approx
8\times 10^{-9}$\,N\,m$^{2}$.  The quantities $\lambda$ and $A$ and
the acceleration of gravity $g$ combine to form the length
$\ell=(A/\lambda g)^{1/3}\approx 5$ cm on which gravity bends a hair
\cite{AudolyPomeau}.

Individual hairs display a range of shapes (Fig.~\ref{fig2}a) which we
have quantified by high-resolution stereoscopic imaging
\cite{details}.  Both the mean squared curvature $\myav{kappa_0(s)^2}$
and the radial excursion $\myav{r_0(s)^2}$ increase with arclength $s$
measured from the top of the switch (Fig.~\ref{fig2}b).  Whilst some
of this is undoubtedly due to gravity (recall $\ell\approx 5$\,cm),
the major part is intrinsic, as we have verified by examining inverted
hairs.  This is in part due to the preparation process: after washing
and rinsing, the hairs in a drying ponytail pass through a glass
transition with decreasing humidity \cite{Robbins}, locking in the
intrinsic curvature \cite{WSC10}, which is naturally reduced in the
vicinity of the clamp due to confinement by neighboring
filaments. Although this is something of a complication when it comes
to interpreting the results, we must regard it as an essential feature
of hair switches and ponytails comprised of \emph{real} fibers. For
later reference, the length-wise averages are
$\myav{\kappa_0^2}=0.15\pm0.01$\,cm$^{-2}$ and $\myav{r_0^2} =
0.80\pm0.05$\,cm$^2$.

\begin{figure*}[t]
\begin{center}
\includegraphics*[clip=true,width=2.00\columnwidth]{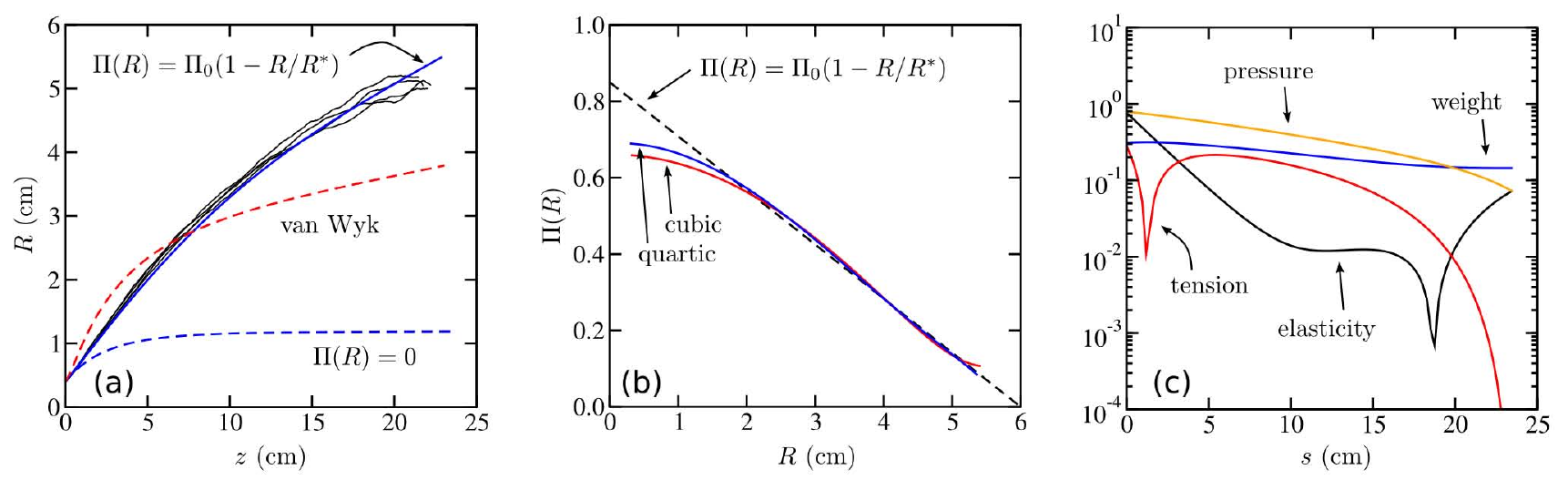}
\end{center}
\vskip -0.5cm
\caption{(color online) Analysis of ponytail shapes.  (a) Measured
  hair switch profiles (thin black lines), compared to the prediction
  of Eq.~\eqref{eq:ndpts} with $\Pi(R)$ as given (solid blue line),
  with $\Pi(R)=0$ (dashed blue line), and with the van Wyk EOS (dashed
  red line) \cite{vanWyk, vwnote} (b) Dimensionless swelling pressure
  $\Pi(R)$ from cubic and quartic fits to the measured profiles, using
  procedure outlined in the text.  (c) The magnitude of the four terms
  in Eq.~\eqref{eq:ndpts} for the calculated profile (solid blue line)
  in (a).}
\label{fig3}
\end{figure*}

Figure \ref{fig3}a shows measured profiles of radius $R(z)$ vs
distance $z$ below the clamp for four separate switches of length
$L\approx 25$\,cm.  Each profile has been obtained from the rotational
average of five images as in Fig.~\ref{fig1}a viewed from angles
$72^{\circ}$ apart. The switch profile shows quite good
reproducibility and is well modeled and explained by the theory we now
describe.  Our starting point for the continuum theory is to introduce
the fiber length density $\rho(\rvec)$ (the number of fibers per unit
area intersecting a plane perpendicular to the fibers) and the mean
fiber tangent vector $\tvec(\rvec)$, the local average of unit vectors
along the fibers.  The latter is a meaningful quantity when the fiber
orientation remains coherent over length scales much larger than the
mean fiber spacing $\rho^{-1/2}$.  Here the fibers are indeed
well-aligned, with $t \equiv |{\bf t}| \approx 1$, unlike in non-woven
fabrics \cite{Kabla}.  In the absence of fiber ends in the bulk these
continuum fields obey a continuity equation ${\bm{\nabla}}\cdot(\rho\,
{\bf t})=0$.  The analogy to the continuity equation of fluid
mechanics mathematizes the remark made by Leonardo at the beginning of
the 16th century, that hair resembles fluid streamlines
\cite{Leonardo}, an observation which has been exploited in more
recent times to aid computer animation \cite{Hadap}. For later use we
also define the local packing fraction $\phi=\pi\rho d^2\!/4$.  We
propose the energy of an axisymmetric fiber bundle is
\begin{equation}
{\cal E}[\rho,{\bf t}]=\int\! d^3\rvec\,
\rho\left(\frac{1}{2}A\kappa^2+\varphi({\bf r})+\myav{u}\right) ~,
\label{eq:e}
\end{equation}
where $\kappa=|(\tvec\cdot\bm{\nabla})\tvec|$ is the curvature field.
The terms in \eqref{eq:e} are the elastic energy of mean curvature,
the external (\eg\ gravitational) potential $\varphi$, and a fiber
confinement energy per unit length $\uav$ that aggregates all terms
involving disorder, such as contacts and natural curvatures.  Without
axisymmetry, \eqref{eq:e} should include terms arising from the
torsion of $\tvec$.  As in density functional theory \cite{HMbook}, we
suppose that $\uav$ is some \emph{local} function of $\rho$.
Minimization of \eqref{eq:e} provides a variational principle for the
bundle shape and the distribution of fibers.  When recast as
mechanical force balance we make contact with the EOS, and identify
$P(\rho)=\rho^2 d\uav/d\rho$ as the pressure.

To address the specific problem of ponytail shapes we now introduce
models which allow for largely analytical calculations.  With
axisymmetry, an integrated form of the continuity equation is $2\pi
r\rho\sin\theta=-\partial n/\partial z$ and $2\pi
r\rho\cos\theta=\partial n/\partial r$, where $n(r, z)$ is the number
of fibers within a radius $r$ at depth $z$ and $\theta$ is the angle
the tangent vector makes to the vertical. ($n(r,z)$ plays the role of
the stream function in fluid mechanics.)  We insert this into
Eq.~\eqref{eq:e}, and use a trial {\it uniform} radial density
function with {\it self-similar} form, $n(r,z)=N[r/R(z)]^2$ where $N$
is the total number of fibers and $R(z)$ is the ponytail radius
(Fig.~\ref{fig1}a).  In practice it is more convenient to use $R(s)$,
where $s(z)$ is the arclength from the clamp.  If $L$ is the
fully-extended hair length and $\varphi=\lambda gz$ the gravitational
potential energy, then to second order in $R_s\equiv dR/ds$,
neglecting a small splay term, one finds
\begin{equation}
{\cal E}= N\!\int_0^L\!\!ds\,\left[\frac{1}{2}\tilde{A}R_{ss}^2+
\frac{1}{2} \tilde{\lambda} g(L-s)R_s^2+\uav\right]~,
\label{eq:c6}
\end{equation}
where $\uav$ depends on $\rhobar=N/(\pi R^2)$.  The problem is now
mapped to an equivalent single fiber hanging under gravity in the
presence of a radial force field derived from $\uav$.  The uniform
distribution in the trial density function yields renormalized
material properties $\tilde{A}=A/2$ and $\tilde{\lambda}=\lambda/2$.
Minimizing Eq.~\eqref{eq:c6} leads to
\begin{equation}
\ell^3 R_{ssss} - (L-s) R_{ss} + R_s - \Pi(R) = 0\label{eq:ndpts}
\end{equation}
where $\Pi(R)={4\ell^3P}/{A\overline\rho R} = -
({2\ell^3}/{A})\,{d\uav}/{dR}$.  We term this the \emph{ponytail shape
  equation}.  It describes a force balance on a length element of the
notionally equivalent single fiber as the sum of four dimensionless
terms which are, respectively, an elastic restoring force, a `string
tension' contribution, a weight term, and a radial swelling force
corresponding to a pressure gradient $P/R$ per unit fiber density.
The ratio ${\rm Ra}\equiv L/\ell$ we shall term the \emph{Rapunzel
  number}, since it is a dimensionless measure of the ponytail length.
When the ponytail hangs from a circular clamp of radius $R_c$, the
boundary conditions are $R(0)=R_{\rm c}$ and $R_s(0)=\tan\theta_c$
where $\theta_c$ is the `launch' angle of the outermost fibers
emerging from the clamp.  At the free bottom of the ponytail the
boundary conditions are $R_{ss}(L)=R_{sss}(L)=0$. To the order at
which we are working, \eqref{eq:ndpts} is supplemented by $z_s\simeq
1-R_s^2/2$ to give the parametric ponytail shape $(z(s), R(s))$.

Fitting the above theory to the experimental ponytail profiles in
Figure \ref{fig3}a reveals a remarkably simple form for the pressure
$\Pi(R)$.  While the full Eq.~\eqref{eq:ndpts} can in principle be
used to determine $\Pi(R)$ from the profiles, the extraction of
high-order derivatives from such data is notoriously problematic.  We
notice though that, away from the clamp, $R_{ssss}$ is likely to be
subdominant to the other terms in Eq.~\eqref{eq:ndpts} and therefore
we can neglect this elastic term and approximate $\Pi(R)\simeq
R_s-(L-s)R_{ss}$, where the right-hand-side is obtained by a low-order
polynomial fit to the data.  Figure \ref{fig3}b shows that in this
region the EOS is accurately represented by
\begin{equation}
\Pi(R)=\Pi_0(1-R/R^*)~,
\label{eq:c66}
\end{equation}
with $\Pi_0\approx0.85$ and $R^*\approx6$\,cm.  Inserting this into
Eq.~\eqref{eq:ndpts} and now \emph{including} the elastic term
recovers the solid blue line in Fig.~\ref{fig3}a, in excellent
agreement with the data (by contrast the van Wyk EOS simply cannot be
made to fit the data \cite{vwnote}).  In making these calculations we
use $R_{\rm c}\approx4$\,mm and $\theta_c\approx 17^\circ$, obtained
from measurements near the clamp.  The starting radius $R_{\rm c}$
corresponds to $\phi\approx0.95$, consistent with the near close
packing of the fibers, whilst the starting angle $\theta_c$ is
presumably governed by the method of clamping (in our case a rubber
band wrapped several times around the top of the switch).

Figure \ref{fig3}c shows the magnitudes of the terms in
Eq.~\eqref{eq:ndpts}.  In the region near the clamp ($s\lesssim
2$\,cm), elasticity and pressure balance, but for the most part the
dominant balance is between weight and pressure, justifying our claim
that the elastic term is subdominant away from the clamp.  The blue
dashed line in Fig.~\ref{fig3}a is the profile for $\Pi(R)=0$.  Since
${\rm Ra}\approx 5$ is quite large, this shape is dominated by
gravity.  Comparing the dashed and solid blue lines in
Fig.~\ref{fig3}a highlights again the dominant role played by the
swelling pressure in determining the shape.

\begin{figure}[t]
\begin{center}
\includegraphics[clip=true,width=0.95\columnwidth]{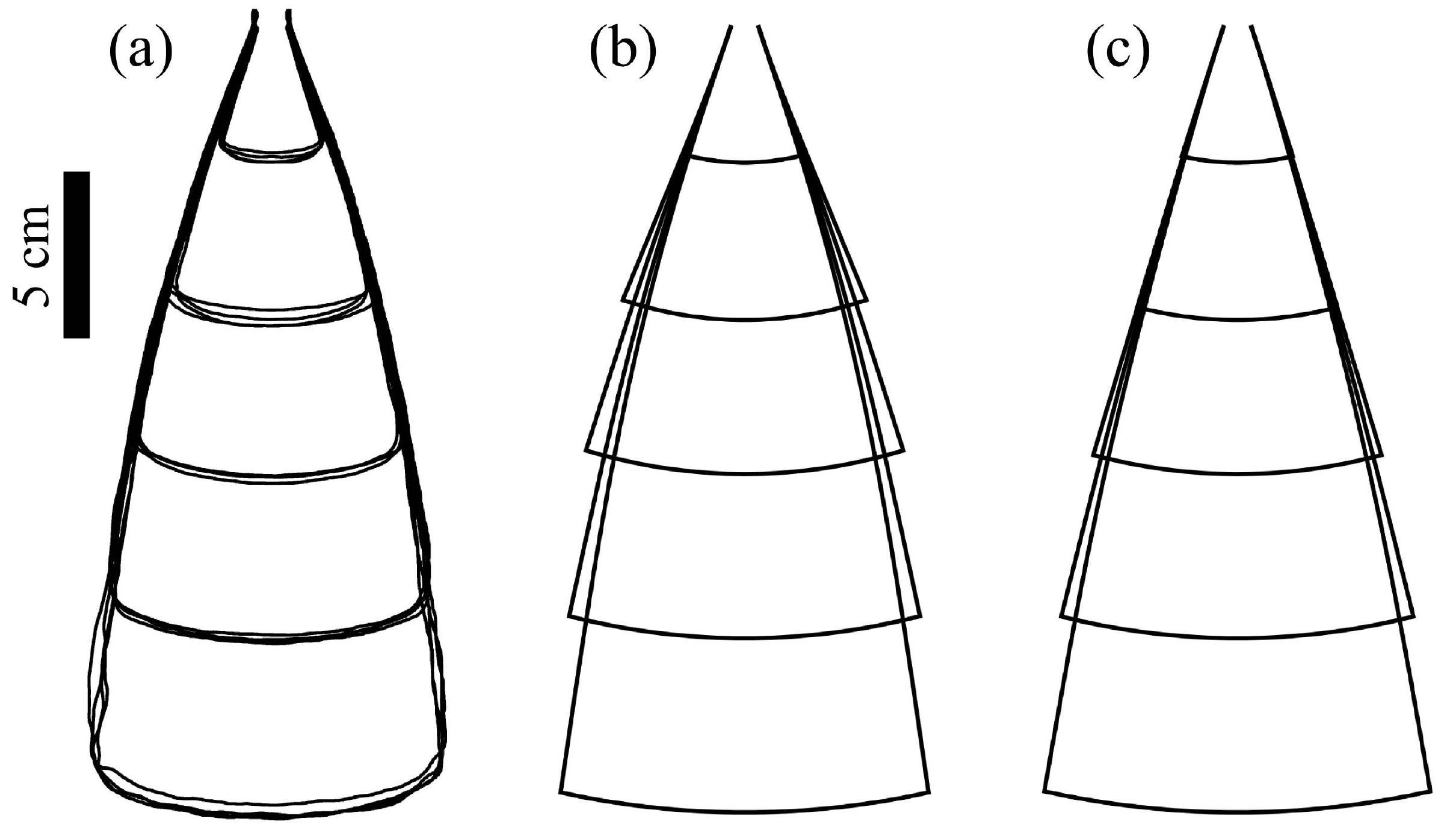}
\end{center}
\vskip -0.3cm
\caption{Trimming a ponytail. (a) Superimposed rotationally-averaged
  outlines of four hair switches, cut down from 25\,cm in steps of
  5\,cm.  (b) Predicted profiles from Eq.~\eqref{eq:ndpts} with
  $\Pi(R)=\Pi_0(1-R/R^*)$.  (c) Predicted profiles with
  $\Pi(R,s)=\Pi_0(1-R/R^*)\times 2s/L^*$ where $L^*=25$\,cm.}
\label{fig4}
\end{figure}

Given $\Pi(R)$, the shape of any ponytail can be predicted.  Thus we
are led to a kind of \latin{experimentum crucis}, shown in
Fig.~\ref{fig4}, in which the predictions of Eq.~\eqref{eq:ndpts} are
compared to the profiles of progressively cut hair switches.  The
agreement is very good.  We observe empirically that the launch angle
$\theta_c$ is remarkably constant at 17$^\circ$, only decreasing to
16$^\circ$ for the shortest hair switch (all calculations used
$\theta_c=17^\circ$).  The calculated profiles show a modest
compaction on increasing length, while the experimental profiles
almost completely collapse on top of one another.  This is not an
effect of plasticity \cite{Kabla} since the switches are compressed in
the cutting stage.  The predicted profiles can similarly collapse
(Fig.~\ref{fig4}c), by allowing for an additional length dependence
reflecting the gradient in the intrinsic fiber properties
(Fig.~\ref{fig2}b).

How are we to interpret the EOS recovered by this analysis?  We
propose that it can be associated with the intrinsic curvatures of the
filaments.  Let us imagine that the effects of collisions with
neighboring fibers can be captured by a \emph{tube model}.
Specifically, consider a helical fiber \cite{helix} of radius $a_0$
confined within a cylinder of radius $a<a_0$, for which $\uav\approx
A\myav{\kappa_0^2}(1-a/a_0)^2/2$ (this is not exact but is quite
accurate).  Matching this to $\uav=(A/2\ell^3)\int_{R}^\infty
\Pi(R)\,dR$, obtained by integrating the expression below
Eq.~\eqref{eq:ndpts}, and inserting our empirical $\Pi(R)$ gives
$a/a_0\approx1-\alpha+\alpha R/R^*$ where $\alpha=\sqrt{\Pi_0
  R^*/2\ell^3\myav{\kappa_0^2}}\approx0.4$.  If we additionally
suppose $a_0^2\approx\myav{r_0^2}$ then for instance at the clamp
($R=R_c$) the confining tube radius $a\approx 6$\,mm.  This seems to
be in reasonable accord with observations, (see
\eg\ Fig.~\ref{fig1}b).  In the tube model, the pressure thus arises
from increasing confinement of the fibers but they are still far from
being completely straightened, even at high compression.  This is
perhaps not surprising considering the role that must eventually be
played by friction.

Of the existing EOS theories, that of van Wyk does not fit our data
(red dashed line, Fig.~\ref{fig3}a), nor does it have an explicit link
to the random curvatures.  The fiber-collision model \cite{Beckrich}
can be extended to three dimensions, but the link to the underlying
statistical properties becomes very unwieldy.  More importantly, in
that model fiber excursions are limited by \emph{nearest-neighbor}
collisions.  This is not necessarily the case in three dimensions, and
in fact is not supported by our data.  Hence the microscopic link
between fiber confinement and packing fraction remains an important
open problem.  Interestingly both the tube model and the collision
model predict that the pressure remains finite on approach to the
close packing limit, in marked contrast to thermal systems of hard
particles.  Thus, a bundle can be collapsed by sufficiently strong
inter-fiber attractions, such as the capillary forces acting on wet
hair \cite{Py07} or a paintbrush.

The program laid out here extends some central paradigms in
statistical physics to the enchanting problem of ponytail shapes.  The
remarkably simple equation of state we have found, along with the
systematic variation of intrinsic curvature along fibers, may open the
way to understanding a wide range of hair and fur geometries.  It is
also of interest to extend the analysis to the \emph{dynamics} of
fiber bundles, epitomized by the `swing' of a ponytail \cite{Kel10},
where the notion of an equivalent single fiber may again prove
fruitful.

We thank A. Avery, M.E. Cates, and A.S. Ferrante for helpful
discussions. This work was supported in part by the Schlumberger Chair
Fund.


\begin{thebibliography}{16}

\bibitem{Robbins} C.R. Robbins, {\it Chemical and Physical Behavior of
  Human Hair} (Springer-Verlag, New York, 2002).

\bibitem{AudolyPomeau} B. Audoly and Y. Pomeau, {\it Elasticity and
  Geometry: From hair curls to the nonlinear response of shells} (OUP,
  Oxford, 2010).

\bibitem{Leonardo}{\it The Notebooks of Leonardo da Vinci},
  ed. J. P. Richter (Dover, London, 1989)

\bibitem{Hadap} S. Hadap and N. Magnenat-Thalmann, in {\it Computer
  Animation and Simulation}, eds. N. Magnenat-Thalmann, D. Thalmann
  and B. Arnaldi (Springer-Verlag, Vienna, 2000); F. Bertails, {\it et
    al.}, ACM Trans. Graphics {\bf 25}, 1180 (2006); R. Bridson and
  C. Batty, Science {\bf 330}, 1756 (2010).

\bibitem{vanWyk} C.M. van Wyk, J. Textile Inst. Trans. {\bf 37}, T285
  (1946).

\bibitem{Carnaby} G. A. Carnaby, R. Postle, and S. de Jong, {\it
  Mechanics of Wool Structures} (Prentice-Hall, New York, 1988).

\bibitem{Beckrich} P. Beckrich, G. Weick, C.M. Marques, and
  T. Charitat, Europhys. Lett.  {\bf 64}, 647 (2003).

\bibitem{IHIP} International Hair Importers \& Products Inc.,
  Glendale, New York.  Switches were washed in mild surfactant
  solution, rinsed and dried before use. Experiments were conducted at
  20$^{\circ}$C and 36--42\% relative humidity.

\bibitem{details} R.E. Goldstein and P.B. Warren, to be
  published.  This algorithm includes a skeletonization method adapted
  from G.J. Stephens, B. Johnson-Kerner, W. Bialek and W.S. Ryu, PLoS
  Comput. Biol. {\bf4}, e1000028 (2008).

\bibitem{WSC10} F.J. Wortmann, M. Stapels and L. Chandra, J. Cosmetic
  Sci. {\bf61}, 31 (2010).

\bibitem{Kabla} A. Kabla and L. Mahadevan, J. R. Soc. Int. {\bf 4}, 99
  (2007).

\bibitem{HMbook} J.-P. Hansen and I.R. McDonald, {\it Theory of Simple
  Liquids}, 3rd ed.  (Academic Press, London, 2006).

\bibitem{vwnote} The van Wyk EOS is essentially $P=kE\phi^3$ but the
  empirical prefactor has to be greatly reduced from the usual value,
  $k\approx0.01$, otherwise the profile `blows up'.  Thus the red
  dashed line in Fig.~\ref{fig3}b has $k=10^{-5}$.

\bibitem{helix} An EOS based on deformed helices in planar confinement
  is in G.A.V. Leaf and W. Oxenham, J. Textile Inst. {\bf 4}, 168
  (1981).

\bibitem{Py07} C. Py, \etal\,, EPL {\bf77}, 44005 (2007).

\bibitem{Kel10} J.B. Keller, SIAM J. Appl. Math. {\bf 70}, 2667 (2010).

\end{thebibliography}
\end{document}